\documentclass[a4paper,11pt]{article}

\usepackage{amsmath,amssymb,bbm}
\usepackage{graphicx}
\usepackage{fancybox}
\usepackage{hyperref}
\hypersetup{
pdftitle={GLSM for exotic five-brane},
pdfauthor={Tetsuji Kimura and Shin Sasaki},
colorlinks={true},
linkcolor={black},
urlcolor={black},
filecolor={black},
citecolor={black}
}
\makeatletter

 \@addtoreset{equation}{section}
\makeatother

\newcounter{Enumerate}

\DeclareFontFamily{U}{rsf}{}
\DeclareFontShape{U}{rsf}{m}{n}{
  <5> <6> rsfs5 <7> <8> <9> rsfs7 <10-> rsfs10}{}
\DeclareMathAlphabet\Scr{U}{rsf}{m}{n}

\usepackage[mathscr]{eucal}

\newcommand{\del}{\partial}

\newcommand{\half}{\frac{1}{2}}

\newcommand{\ls}{\ \ \ \ \ }
\newcommand{\wt}{\widetilde}

\newcommand{\ve}{\varepsilon}
\newcommand{\ol}{\overline}

\newcommand{\bsubeq}{\begin{subequations}}
\newcommand{\esubeq}{\end{subequations}}
\newcommand{\noi}{\noindent}
\newcommand{\mr}{\mathring}

\newcommand{\nn}{\nonumber}
\newcommand{\N}{\mathcal{N}}
\renewcommand{\d}{{\rm d}}
\newcommand{\e}{{\rm e}}
\renewcommand{\i}{{\rm i}}
\newcommand{\w}{\wedge}
\newcommand{\slb}{\scalebox}

\def\+{{+\!\!\!+}} 

\begin{document}
\allowdisplaybreaks{

\thispagestyle{empty}


\begin{flushright}
RUP-13-4
\end{flushright}

\vspace{30mm}

\noi
\slb{2.5}{Gauged Linear Sigma Model}

\vspace{3mm}

\noi
\slb{2.5}{for Exotic Five-brane}

\vspace{7mm}

\noi
{\renewcommand{\arraystretch}{1.7}
\begin{tabular}{cl}
\multicolumn{2}{l}{\slb{1.0}{Tetsuji {\sc Kimura}}}
\\
& \slb{.9}{\renewcommand{\arraystretch}{1.0}
\begin{tabular}{l}
{\sl
Department of Physics and Research Center for Mathematical Physics,
Rikkyo University} 
\\
{\sl Tokyo 171-8501, JAPAN
}
\\
\slb{0.9}{\tt tetsuji \_at\_ rikkyo.ac.jp}
\end{tabular}
}
\\
& \multicolumn{1}{l}{and}
\\
\multicolumn{2}{l}{\slb{1.0}{Shin {\sc Sasaki}}}
\\
& \slb{.9}{\renewcommand{\arraystretch}{1.0}
\begin{tabular}{l}
{\sl
Department of Physics,
Kitasato University}
\\
{\sl Sagamihara 252-0373, JAPAN}
\\
\slb{0.9}{\tt shin-s \_at\_ kitasato-u.ac.jp}
\end{tabular}
}
\end{tabular}
}

\vspace{14mm}


\noindent
\slb{1.1}{\sc Abstract}:
\begin{center}
\slb{.95}{
\begin{minipage}{.95\textwidth}
We study an $\N=(4,4)$ supersymmetric gauged linear sigma model which gives rise to the nonlinear sigma model for multi-centered KK-monopoles.
We find a new T-duality transformation of the model even in the presence of F-terms.
Performing T-duality, 
we find the gauged linear sigma model whose IR limit describes the exotic $5_2^2$-brane with B-field.
\end{minipage}
}
\end{center}

\newpage
\section{Introduction}
\label{sect-introduction}

Applying string duality transformations to familiar objects such as D-branes, 
one often encounters unfamiliar extended objects whose physical properties look exotic.
For example, the mass of the branes
is not strictly well-defined;
the spacetime metric is not single-valued; 
which reflect the exotic nature of these objects.
One such example is known as exotic branes.
The exotic branes appear in lower-dimensional theories in string compactifications.
Some of them are easily constructed via torus compactifications of string theory and M-theory \cite{Obers:1998fb}.

Such exotic branes appear in various situations in string theory.
For instance, in the flux compactification scenarios,
there exist non-geometric backgrounds.
They emerge via string duality transformations from geometric backgrounds.
One typical space is T-fold \cite{Hull:2004in}.
An exotic five-brane, named $5_2^2$-brane, is a concrete realization of T-fold.
Furthermore, the exotic branes should plays an significant role in the black hole quantum mechanics \cite{deBoer:2010ud, deBoer:2012ma}.
In lower-dimensional supergravity theories, deformation parameters can also be interpreted as the contribution of such branes \cite{Bergshoeff:2012pm}.

It is known that the $5_2^2$-brane is T-dual of the Kaluza-Klein (KK) monopoles in type II string theory.
Applying the Buscher rule \cite{Buscher:1987sk} to the geometry of the KK-monopoles, one finds the $5_2^2$-brane geometry.
The four-dimensional transverse space of the KK-monopoles is the Taub-NUT geometry. 
On the other hand, the four-dimensional transverse space of the $5_2^2$-brane is expressed as the $T^2$-fibration\footnote{Because of the $T^2$-compactification, the $5_2^2$-brane is regarded as one of the defect branes \cite{Bergshoeff:2011se}. In this paper, we also utilize the terminology ``defect branes'' to express other codimension two five-branes.} 
over ${\mathbb R}^2$ 
(for the explicit expression, see appendix \ref{sect-fivebrane-sols}). 
Moving around the $5_2^2$-brane on the base space ${\mathbb R}^2$, 
the size of the fibred two-torus does not come back to itself. 
Even though such an unusual feature, the $5_2^2$-brane is also a solution of supergravity.

Although the properties of the $5_2^2$-branes in the supergravity picture have been discussed \cite{deBoer:2010ud, Kikuchi:2012za, deBoer:2012ma},
the string worldsheet description is still less understood.
The worldsheet descriptions of the H-monopoles and KK-monopoles are well investigated in the language of the supersymmetric gauge theory, called the gauged linear sigma model (GLSM) \cite{W93, Mirror03}. 
The GLSM is a powerful tool to
study non-perturbative corrections in string theory \cite{Morrison:1994fr}.
Indeed, the stringy winding corrections of the H- and KK-monopole geometries are studied through an examination of the instanton effects in the GLSMs \cite{Tong:2002rq, Harvey:2005ab, Okuyama:2005gx}.

In this paper, we investigate the GLSM formulation of the exotic $5_2^2$-brane.
It is difficult to construct the GLSM for the $5_2^2$-brane in a straight manner. 
This is because the $5_2^2$-brane is an object of codimension two,
whose description is ill-defined in the asymptotic region.
Thus we have to introduce the renormalization scale. 
However, this dimensionful parameter cannot be involved into the GLSM.
Then we start from the GLSM for multi-centered KK-monopoles of codimension three. 
Arraying an infinite number of the KK-monopoles along one compactified direction, we construct a single defect five-brane of codimension two.
Then, performing the T-duality transformation, 
we find the sigma model whose target space is the exotic $5_2^2$-brane.
This is exactly the same procedure to construct the $5_2^2$-brane in supergravity.
We note that F-terms in the GLSM often prevent the existence of isometries in the target space geometry of the low-energy effective theory.
In order to perform the duality transformation \cite{Rocek:1991ps},
we rewrite the F-terms to D-terms with a trick. 
In addition, we also have to realize the exotic feature of the $5_2^2$-brane, i.e., the ``non-geometric'' structure.
We will elaborate on this at the final step of the analysis.

The organization of this paper is as follows:
In section \ref{sect-Hmono} we briefly review the setup of the GLSM and the duality transformation technique. 
In section \ref{sect-multiNS5KKM-522} we generalize the GLSM in order to apply a further duality transformation.
We carefully analyze the structure of supersymmetric vacua.
In the IR limit, we obtain the nonlinear sigma model which contains the non-dynamical field. 
This can be interpreted as the non-geometric coordinate in the viewpoint of the target space geometry.
Integrating it out, 
we finally obtain the correct sigma model whose target space is the exotic $5_2^2$-brane.
Section \ref{sect-conclusion} is devoted to conclusion and discussions.
In appendix \ref{sect-fivebrane-sols} we exhibit a few examples of five-brane configurations.
In appendix \ref{sect-another5} we briefly mention the GLSM which differs from the ones for KK-monopoles.

\section{Review of GLSMs and T-duality}
\label{sect-Hmono}

In this section we introduce the supersymmetric GLSMs of our interest. 
We also briefly review the duality transformation technique in the framework of the superfield formalism.
Our approach in this section is based on the works \cite{Tong:2002rq, Harvey:2005ab, Okuyama:2005gx}.

\subsection{Multi-centered H-monopoles}

We begin with the two-dimensional $\N=(4,4)$ supersymmetric $U(1)^k$ gauge theory coupled to $k$ charged hypermultiplets and one neutral twisted hypermultiplet in the presence of $k$ complex Fayet-Iliopoulos (FI) parameters $(s_a, t_a)$ \cite{Okuyama:2005gx}:
\begin{align}
\Scr{L}_1 \ &= \ 
\sum_{a=1}^k \int \d^4 \theta \, \Big\{ 
\frac{1}{e_a^2} 
\Big( - \ol{\Sigma}{}_a \Sigma_a + \ol{\Phi}{}_a \Phi_a \Big)
+ \ol{Q}{}_a \, \e^{-2 V_a} Q_a
+ \ol{\wt{Q}}{}_a \, \e^{+2 V_a} \wt{Q}_a
\Big\}
+ \int \d^4 \theta \, 
\frac{1}{g^2} 
\Big( - \ol{\Theta} \Theta + \ol{\Psi} \Psi \Big)
\nn \\
\ & \ \ \ \ 
+ \sum_{a=1}^k \Big\{
\sqrt{2} \int \d^2 \theta \, \Big( \wt{Q}_a \Phi_a Q_a + (s_a - \Psi) \Phi_a
\Big)
+ \text{(h.c.)} 
\Big\}
\nn \\
\ & \ \ \ \ 
+ \sum_{a=1}^k \Big\{ 
\sqrt{2} \int \d^2 \wt{\theta} \, \big( t_a - \Theta \big) \Sigma_a
+ \text{(h.c.)}
\Big\}
\, . \label{multi-GLSM51} 
\end{align}
Here we have two coupling constants.
One is the dimensionful gauge coupling constant $e_a$, 
while the other is the dimensionless sigma model coupling constant $g$.
The constituents of the $\N=(4,4)$ supermultiplets are described in terms of the $\N=(2,2)$ supermultiplets in the following way:
\begin{align*}
\text{$\N=(4,4)$ vector multiplets} &: \ \ \ 
\left\{
\begin{array}{rl}
\Sigma_a :& \text{$\N=(2,2)$ twisted chiral multiplets} \\
\Phi_a :& \text{$\N=(2,2)$ chiral multiplets} 
\end{array}
\right.
\\
\text{$\N=(4,4)$ charged hypermultiplets} &: \ \ \ 
\left\{
\begin{array}{rl}
Q_a :& \text{$\N=(2,2)$ chiral multiplets with charge $-1$} \\
\wt{Q}_a :& \text{$\N=(2,2)$ chiral multiplets with charge $+1$} 
\end{array}
\right.
\\
\text{$\N=(4,4)$ twisted hypermultiplet} &: \ \ \ 
\left\{
\begin{array}{rl}
\Psi :& \text{$\N=(2,2)$ chiral multiplet} \\
\Theta :& \text{$\N=(2,2)$ twisted chiral multiplet}
\end{array}
\right.
\end{align*}
The terms which do not involve derivatives in the component expansion of these superfields are
(for a detailed introduction to $\N=(2,2)$ theories, see \cite{W93, Mirror03}) 
\bsubeq
\begin{align}
V_a \ &= \ 
\theta^+ \ol{\theta}{}^+ (A_{0,a} + A_{1,a})
+ \theta^- \ol{\theta}{}^- (A_{0,a} - A_{1,a})
- \sqrt{2} \, \theta^- \ol{\theta}{}^+ \sigma_a
- \sqrt{2} \, \theta^+ \ol{\theta}{}^- \ol{\sigma}{}_a
\nn \\
\ & \ \ \ \ 
- 2 \i \, \theta^+ \theta^-
\big( \ol{\theta}{}^+ \ol{\lambda}{}_{+,a} 
+ \ol{\theta}{}^- \ol{\lambda}{}_{-,a} \big)
+ 2 \i \, \ol{\theta}{}^+ \ol{\theta}{}^-
\big( \theta^+ \lambda_{+,a} + \theta^- \lambda_{-,a} \big)
- 2 \, \theta^+ \theta^- \ol{\theta}{}^+ \ol{\theta}{}^- D_{V,a}
\, , \\
\Sigma_a 
\ &= \ 
\frac{1}{\sqrt{2}} \ol{D}{}_+ D_- V_a
\, , \\
\Phi \ &= \ 
\phi 
+ \i \sqrt{2} \, \theta^+ \wt{\lambda}_+ 
+ \i \sqrt{2} \, \theta^- \wt{\lambda}_-
+ 2 \i \, \theta^+ \theta^- D_{\Phi} 
+ \ldots
\, , \\
Q_a \ &= \ 
q_a 
+ \i \sqrt{2} \, \theta^+ \psi_{+,a} 
+ \i \sqrt{2} \, \theta^- \psi_{-,a}
+ 2 \i \, \theta^+ \theta^- F_a
+ \ldots
\, , \\
\wt{Q}_a \ &= \ 
\wt{q}_a
+ \i \sqrt{2} \, \theta^+ \wt{\psi}_{+,a}
+ \i \sqrt{2} \, \theta^- \wt{\psi}_{-,a}
+ 2 \i \, \theta^+ \theta^- \wt{F}_a
+ \ldots
\, , \\
\Psi \ &= \ 
\frac{1}{\sqrt{2}} (r^1 + \i r^2)
+ \i \sqrt{2} \, \theta^+ \chi_+
+ \i \sqrt{2} \, \theta^- \chi_-
+ 2 \i \, \theta^+ \theta^- G
+ \ldots
\, , \\
\Theta \ &= \ 
\frac{1}{\sqrt{2}} (r^3 + \i \vartheta)
+ \i \sqrt{2} \, \theta^+ \ol{\wt{\chi}}{}_+
+ \i \sqrt{2} \, \ol{\theta}{}^- \wt{\chi}_-
+ 2 \i \, \theta^+ \ol{\theta}{}^- \wt{G}
+ \ldots
\, , 
\end{align}
\esubeq
where the vector superfields $V_a$ are in the Wess-Zumino gauge.
The symbol ``$\ldots$'' implies the derivative terms governed by 
the covariant derivatives
\begin{alignat}{2}
D_{\pm} \ &= \ 
\frac{\del}{\del \theta^{\pm}}
- \i \, \ol{\theta}{}^{\pm} (\del_0 \pm \del_1)
\, , &\ls
\ol{D}{}_{\pm} \ &= \ 
- \frac{\del}{\del \ol{\theta}{}^{\pm}}
+ \i \, \theta^{\pm} (\del_0 \pm \del_1)
\, . 
\end{alignat}
The complex FI parameters $(s_a, t_a)$ are decomposed into the real and imaginary parts:
\begin{align}
t_a \ = \ 
\frac{1}{\sqrt{2}} (t_{1,a} + \i \, t_{2,a})
\, , \ls
s_a \ = \ 
\frac{1}{\sqrt{2}} (s_{1,a} + \i \, s_{2,a})
\, . 
\end{align}

In the IR limit $e_a \to \infty$, 
we obtain the effective theory. 
This is given as the nonlinear sigma model whose target space gives rise to the multi-centered H-monopoles with B-field.
The values of the FI parameters $(s_a, t_a)$ represent the centers of the H-monopoles in the four-dimensional transverse directions \cite{Okuyama:2005gx}.

\subsection{Multi-centered KK-monopoles}

Here we exhibit the duality transformation technique which give rises to the T-duality transformation in the viewpoint of the target space geometry of the effective theory.

First, we focus on the terms containing the twisted chiral superfield $\Theta$ in (\ref{multi-GLSM51}) and rewrite it as follows:
\bsubeq
\begin{align}
\Scr{L}_{\Theta} \ &= \ 
\int \d^4 \theta \, \Big( - \frac{1}{g^2} \ol{\Theta} \Theta \Big)
+ \sum_{a=1}^k \Big\{ \sqrt{2} \int \d^2 \wt{\theta} \, 
\big( - \Theta \big) \Sigma_a 
+ \text{(h.c.)}
\Big\}
\nn \\
\ &= \ 
\int \d^4 \theta \, \Big\{
- \frac{1}{2 g^2} B^2
- 2 B \sum_{a=1}^k V_a 
- (\Gamma + \ol{\Gamma}) B
\Big\}
- \sqrt{2} \, \ve^{mn} \sum_{a=1}^k \del_m (\vartheta A_{a,n})
\, , \label{dual1-GLSM5}
\end{align}
\esubeq
where we introduced an auxiliary real superfield $B$ and an auxiliary chiral superfield $\Gamma$.
Integrating out the auxiliary superfield $\Gamma$, we find a constraint on the real superfield $B$ in such a way as
\begin{align}
\ol{D}{}_{+} \ol{D}{}_- B \ = \ 0 \ = \ D_{+} D_- B
\ \ \ \to \ \ \ 
B \ = \ \Theta + \ol{\Theta}
\, . \label{B2Theta}
\end{align}
Plugging this solution into the second line of (\ref{dual1-GLSM5}), 
we can go back to the first line.
Instead of this, we obtain another constraint when we integrate out the real superfield $B$:
\begin{align}
\frac{1}{g^2} B \ &= \ 
- (\Gamma + \ol{\Gamma}) 
- 2 \sum_{a=1}^k V_a
\, . \label{B2GammaV}
\end{align}
Applying this constraint to (\ref{dual1-GLSM5}),
we obtain the following gauge theory:
\begin{align}
\Scr{L}_2 \ &= \ 
\sum_{a=1}^k \int \d^4 \theta \, \Big\{ 
\frac{1}{e_a^2} 
\Big( - \ol{\Sigma}{}_a \Sigma_a + \ol{\Phi}{}_a \Phi_a \Big)
+ \ol{Q}{}_a \, \e^{-2 V_a} Q_a
+ \ol{\wt{Q}}{}_a \, \e^{+2 V_a} \wt{Q}_a
\Big\}
\nn \\
\ & \ \ \ \
+ \int \d^4 \theta \, 
\frac{g^2}{2} 
\Big( \Gamma + \ol{\Gamma} + 2 \sum_{a=1}^k V_a \Big)^2
- \sqrt{2} \, \ve^{mn} \sum_{a=1}^k \del_m (\vartheta A_{a,n})
+ \int \d^4 \theta \, 
\frac{1}{g^2} \ol{\Psi} \Psi
\nn \\
\ & \ \ \ \ 
+ \sum_{a=1}^k \Big\{
\sqrt{2} \int \d^2 \theta \, \Big( \wt{Q}_a \Phi_a Q_a + (s_a - \Psi) \Phi_a
\Big)
+ \text{(h.c.)} 
\Big\}
\nn \\
\ & \ \ \ \ 
+ \sum_{a=1}^k \Big\{ 
\sqrt{2} \int \d^2 \wt{\theta} \, t_a \Sigma_a
+ \text{(h.c.)}
\Big\}
\, . \label{multi-GLSM52}
\end{align}
These two theories are equivalent to each other 
since $\Theta$ and $\Gamma$ are related through the equations (\ref{B2Theta}) and (\ref{B2GammaV}):
\begin{align}
\Theta + \ol{\Theta}
\ &= \ 
- g^2 (\Gamma + \ol{\Gamma}) - 2 g^2 \sum_{a=1}^k V_a
\, . \label{Theta-Gamma-V-GLSM5}
\end{align}
Some of the component fields of $\Gamma$ 
are assigned by the ones of $\Theta$ and $V_a$ via (\ref{Theta-Gamma-V-GLSM5}):
\bsubeq \label{Theta-Gamma-V-comp-GLSM5}
\begin{gather}
\Gamma \ = \ 
\frac{1}{\sqrt{2}} \Big( - \frac{r^3}{g^2} + \i \, \gamma^4 \Big)
- \frac{\i}{g^2} \sqrt{2} \, \theta^+ \ol{\wt{\chi}}{}_{+}
- \frac{\i}{g^2} \sqrt{2} \, \theta^- \ol{\wt{\chi}}{}_{-}
+ 2 \i \, \theta^+ \theta^- G_{\Gamma}
+ \ldots
\, ,
\end{gather}
whilst the component field $\gamma^4$ of $\Gamma$ is related to $\vartheta$ of $\Theta$ only through the following way:
\begin{gather}
\pm (\del_0 \pm \del_1) \vartheta
\ = \ 
- g^2 \Big\{
(\del_0 \pm \del_1) \gamma^4
+ \sqrt{2} \sum_{a=1}^k (A_{0,a} \pm A_{1,a})
\Big\}
\, .
\end{gather}
\esubeq
This is the duality transformation procedure in the superfield formalism \cite{Rocek:1991ps}.

When we take the IR limit $e_a \to \infty$ of the theory $\Scr{L}_2$,
we obtain the nonlinear sigma model for multi-centered KK-monopoles \cite{Okuyama:2005gx}, which is also interpreted as the T-dualized description of the multi-centered H-monopoles via the Buscher rule \cite{Buscher:1987sk}.
We note that the topological term in the Lagrangian (\ref{multi-GLSM52}) contains the real scalar field $\vartheta$ which is not the coordinate of the KK-monopole geometry.
Even though this is no longer dynamical, this plays an important role in the quantum corrections of the system \cite{Harvey:2005ab}.

\section{GLSM for exotic five-brane}
\label{sect-multiNS5KKM-522}

In this section we generalize the GLSM for the multi-centered KK-monopoles (\ref{multi-GLSM52}).
First we introduce a trick in order to apply the duality transformation to the chiral superfield $\Psi$.
After that, we investigate the structure of vacua and the effective theory.
Finally we find the nonlinear sigma model whose target space is nothing but the exotic $5_2^2$-brane geometry with B-field.

\subsection{A further dualized Lagrangian}

In the previous section,
we have already dualized the twisted chiral superfield $\Theta$ of the $\N=(4,4)$ neutral twisted hypermultiplet to the chiral superfield $\Gamma$ (\ref{B2GammaV}).
In the dualization procedure, 
we first rewrote the twisted F-term to the D-term (\ref{dual1-GLSM5}).
This is because, in the usual sense, 
the existence of (twisted) F-terms would prevent the existence of 
shift symmetries. They are mapped to
isometries on the target space geometry of the nonlinear sigma model.
In the same analogy, let us first rewrite the F-term $\Psi \Phi_a$ in (\ref{multi-GLSM52}) to D-terms. 
Notice that the chiral superfield $\Phi_a$ in the $\N=(4,4)$ vector multiplet can be described in favor of an unconstrained complex superfield $C_a$ 
\begin{align}
\Phi_a \ = \ \ol{D}{}_+ \ol{D}{}_- C_a
\, . \label{Phi2C}
\end{align}
We extract the terms involving $\Psi$ in (\ref{multi-GLSM52}) and rewrite the F-term to the D-terms: 
\begin{align}
\Scr{L}_{\Psi} \ &= \ 
\int \d^4 \theta \, 
\frac{1}{g^2} \ol{\Psi} \Psi 
+ \sum_{a=1}^k \Big\{
\sqrt{2} \int \d^2 \theta \, (- \Psi) \Phi_a 
+ \text{(h.c.)}
\Big\}
\nn \\
\ &= \
\int \d^4 \theta \, \Big\{
\frac{c}{g^2} (\Psi + \ol{\Psi})^2
- \sqrt{2} (\Psi + \ol{\Psi}) \sum_{a=1}^k (C_a + \ol{C}{}_a)
\Big\}
\nn \\
\ & \ \ \ \ 
+ \int \d^4 \theta \, \Big\{
\frac{2 c - 1}{2 g^2} (\Psi - \ol{\Psi})^2
- \sqrt{2} (\Psi - \ol{\Psi}) \sum_{a=1}^k (C_a - \ol{C}{}_a)
\Big\}
\, . \label{before-Psidual}
\end{align}
Here $c$ is an arbitrary constant.
We replace $\Psi \pm \ol{\Psi}$ to auxiliary fields as follows:
\begin{align}
\Scr{L}_{RSX \Xi} \ &= \ 
\int \d^4 \theta \, \Big\{
\frac{c}{g^2} R^2
- \sqrt{2} R \sum_{a=1}^k (C_a + \ol{C}{}_a)
+ R (\Xi + \ol{\Xi})
+ R (X + \ol{X})
\Big\}  
\nn \\
\ & \ \ \ \ 
+ \int \d^4 \theta \, \Big\{
\frac{2 c - 1}{2 g^2} (\i S)^2
- \sqrt{2} (\i S) \sum_{a=1}^k (C_a - \ol{C}{}_a)
+ (\i S) (\wt{\Xi} - \ol{\wt{\Xi}}) 
+ (\i S) (X - \ol{X})
\Big\}  
\, , \label{dual2-GLSM5}
\end{align}
where $R$ and $S$ are auxiliary real superfields, 
$\Xi$ and $\wt{\Xi}$ are auxiliary twisted chiral superfields,
and $X$ is an auxiliary chiral superfield.
Integrating out $\Xi$ and $\wt{\Xi}$, we find
\bsubeq \label{R2Psi}
\begin{alignat}{3}
\ol{D}{}_+ D_- R \ &= \ 0 \ = \ D_+ \ol{D}{}_- R
\ \ \ &&\to \ \ \ 
&
R \ &= \ \Psi_1 + \ol{\Psi}{}_1
\, , \label{EOM-Xi1} \\
\ol{D}{}_+ D_- (\i S) \ &= \ 0 \ = \ D_+ \ol{D}{}_- (\i S)
\ \ \ &&\to \ \ \ 
&
\i S \ &= \ \Psi_2 - \ol{\Psi}{}_2
\, . \label{EOM-Xi2}
\end{alignat}
\esubeq
Here $\Psi_1$ and $\Psi_2$ are chiral superfields.
Under this constraint we further integrate out the auxiliary field $X$:
\bsubeq \label{EOM-X}
\begin{align}
0 \ &= \ 
\ol{D}{}_+ \ol{D}{}_- (R + \i S) 
\ = \ 
\ol{D}{}_+ \ol{D}{}_- (\ol{\Psi}{}_1 - \ol{\Psi}{}_2)
\, , \\
0 \ &= \ 
D_+ D_- (R - \i S)
\ = \ 
D_+ D_- (\Psi_1 - \Psi_2)
\, .
\end{align}
The only one solution which satisfies the above equations is 
\begin{align}
\Psi_1 \ &= \ \Psi_2
\, . \label{Psi1=Psi2}
\end{align}
\esubeq
Plugging this into (\ref{dual2-GLSM5}), we obtain the same form as (\ref{before-Psidual}). Here we can regard that $\Psi_1$ is nothing but the $\Psi$ in the Lagrangian (\ref{before-Psidual}).

Here we go back to (\ref{dual2-GLSM5}) and consider another configuration.
We integrate out the auxiliary twisted chiral superfield $\wt{\Xi}$
and the auxiliary real superfield $R$.
Each solution is given by (\ref{EOM-Xi2}) and
\begin{align}
0 \ &= \ 
\frac{2c}{g^2} R - \sqrt{2} \sum_{a=1}^k (C_a + \ol{C}{}_a)
+ (\Xi + \ol{\Xi})
+ (X + \ol{X})
\nn \\
\ &= \ 
\frac{2c}{g^2} R - \sqrt{2} \sum_{a=1}^k (C'_a + \ol{C}{}'_a)
+ (\Xi + \ol{\Xi})
\, . \label{R2XiC}
\end{align}
Here the unconstrained complex superfield $C_a$ is changed to $C'_a$ 
since, at least in the classical level, 
the auxiliary chiral superfield $X$ can be absorbed into it
by the definition $\Phi_a = \ol{D}{}_+ \ol{D}{}_- C_a$.
Plugging (\ref{EOM-Xi2}) and (\ref{R2XiC}) into (\ref{dual2-GLSM5}),
we obtain another Lagrangian dual to (\ref{multi-GLSM52}):
\begin{align}
\Scr{L}_3 \ &= \ 
\sum_{a=1}^k \int \d^4 \theta \, \Big\{ 
\frac{1}{e_a^2} 
\Big( - \ol{\Sigma}{}_a \Sigma_a + \ol{\Phi}{}_a \Phi_a \Big)
+ \ol{Q}{}_a \, \e^{-2 V_a} Q_a
+ \ol{\wt{Q}}{}_a \, \e^{+2 V_a} \wt{Q}_a
\Big\}
\nn \\
\ & \ \ \ \ 
+ \int \d^4 \theta \, 
\frac{g^2}{2} 
\Big( \Gamma + \ol{\Gamma} + 2 \sum_{a=1}^k V_a \Big)^2
- \sqrt{2} \, \ve^{mn} \sum_{a=1}^k \del_m (\vartheta A_{a,n})
\nn \\
\ & \ \ \ \ 
+ \int \d^4 \theta \, \Big\{
- \frac{g^2}{2} \Big( \Xi + \ol{\Xi} - \sqrt{2} \sum_{a=1}^k (C'_a + \ol{C}{}'_a) \Big)^2
- \sqrt{2} (\Psi_2 - \ol{\Psi}_2) \sum_{a=1}^k (C'_a - \ol{C}{}'_a)
\Big\}  
\nn \\
\ & \ \ \ \ 
+ \sum_{a=1}^k \Big\{
\sqrt{2} \int \d^2 \theta \, \big( \wt{Q}_a \Phi_a Q_a + s_a \, \Phi_a \big)
+ \text{(h.c.)} 
\Big\}
+ \sum_{a=1}^k \Big\{ 
\sqrt{2} \int \d^2 \wt{\theta} \, t_a \Sigma_a
+ \text{(h.c.)}
\Big\}
\, . \label{multi-GLSM54}
\end{align}
Here we should choose $c = \half$ in order to remove the kinetic term of $\Psi_2$ from the dualized system, because
$\Psi_2$ is subject to the following relation via (\ref{R2Psi}) and (\ref{R2XiC}):
\bsubeq \label{Psi-Xi-C-GLSM5}
\begin{align}
R \ = \ 
\Psi_1 + \ol{\Psi}{}_1
\ &= \ 
\Psi + \ol{\Psi}
\ = \ 
- g^2 (\Xi + \ol{\Xi})
+ \sqrt{2} \, g^2 \sum_{a=1}^k (C'_a + \ol{C}{}'_a)
\, , \\
\i S \ = \ 
\Psi_2 - \ol{\Psi}{}_2
\ &= \ 
\Psi - \ol{\Psi}
\, .
\end{align}
\esubeq
In order to symbolize the relation (\ref{Psi-Xi-C-GLSM5}), 
we re-express the chiral superfield $\Psi_2$ to $\mr{\Psi}$ in later discussions.
Without any confusion, we also rewrite $C'_a$ to $C_a$ because the auxiliary chiral superfield $X$ is completely absorbed and does not explicitly appear in the dualized system (\ref{multi-GLSM54}). 

It seems strange that $\mr{\Psi} - \ol{\mr{\Psi}}$ exists in the dualized Lagrangian (\ref{multi-GLSM54}) because this contains not only $\del_m r^2$ but also $r^2$. This would prevent the shift symmetry $r^2 \to r^2 + \alpha$, where $\alpha$ is arbitrary. The shift symmetry is mapped to the isometry along the $r^2$-direction on the target space geometry of the low energy effective theory.
However, in the later discussion, 
we will find that this term is inevitable to complete the correct T-duality transformation.

There are remarks on the auxiliary fields in (\ref{dual2-GLSM5}):
\begin{itemize}
\item If $X$ is not introduced, one cannot find the coincidence (\ref{Psi1=Psi2}), which is essential to go back to the GLSM for KK-monopoles $\Scr{L}_2$ (\ref{multi-GLSM52}).

\item If one integrates out the pair $(S, \Xi)$ instead of the integrating-out of the pair $(R, \wt{\Xi})$, one obtains another dualized Lagrangian in which the real part of $\Psi$ is T-dualized.
This is intrinsically the same as (\ref{multi-GLSM54}), in which, we will discuss in the next subsection, the imaginary part of $\Psi$ is T-dualized.

\item If the pair $(R, S)$ is integrate out,
all the component fields in $\Psi$ is mapped to $\Xi$.
This does not imply the correct T-duality transformation from the viewpoint of the target space geometry of the theory in the IR limit.
\end{itemize}

Applying the duality transformation (\ref{dual2-GLSM5}) to the original Lagrangian $\Scr{L}_1$ (\ref{multi-GLSM51}), 
we find another Lagrangian $\wt{\Scr{L}}_2$ (see appendix \ref{sect-another5}).
In this paper we do not investigate this system since this will not describe the exotic five-brane which we want.

\subsection{Component fields}
\label{sect-dual-comp-GLSM5}

In order to understand the duality relation (\ref{Psi-Xi-C-GLSM5}) 
in the language of component fields,
let us expand the superfields as follows:
\bsubeq
\begin{align}
\Xi \ &= \ 
\frac{1}{\sqrt{2}} (y^1 + \i \, y^2)
+ \i \sqrt{2} \, \theta^+ \ol{\xi}{}_+
+ \i \sqrt{2} \, \ol{\theta}{}^- \xi_-
+ 2 \i \, \theta^+ \ol{\theta}{}^- G_{\Xi}
+ \ldots
\, , \\
C_a \ &= \ 
\phi_{c,a} + \i \sqrt{2} \, \theta^+ \psi_{c+,a} 
+ \i \sqrt{2} \, \theta^- \psi_{c-,a} 
+ \i \sqrt{2} \, \ol{\theta}{}^+ \chi_{c+,a} 
+ \i \sqrt{2} \, \ol{\theta}{}^- \chi_{c-,a}
\nn \\
\ & \ \ \ \ 
+ 2 \i \, \theta^+ \theta^- F_{c,a} 
+ 2 \i \, \ol{\theta}{}^+ \ol{\theta}{}^- M_{c,a}
+ 2 \i \, \theta^+ \ol{\theta}{}^- G_{c,a} 
+ 2 \i \, \ol{\theta}{}^+ \theta^- N_{c,a}
+ \theta^- \ol{\theta}{}^- A_{c=,a}
+ \theta^+ \ol{\theta}{}^+ B_{c\+,a}
\nn \\
\ & \ \ \ \ 
- 2 \i \, \theta^+ \theta^- \ol{\theta}{}^+ \zeta_{c+,a}
- 2 \i \, \theta^+ \theta^- \ol{\theta}{}^- \zeta_{c-,a}
+ 2 \i \, \ol{\theta}{}^+ \ol{\theta}{}^- \theta^+ \lambda_{c+,a}
+ 2 \i \, \ol{\theta}{}^+ \ol{\theta}{}^- \theta^- \lambda_{c-,a}
- 2 \theta^+ \theta^- \ol{\theta}{}^+ \ol{\theta}{}^- D_{c,a}
\, .
\end{align}
\esubeq
Because of the definition $\Phi_a = \ol{D}{}_+ \ol{D}{}_- C_a$, 
we find the relations among the component fields:
\bsubeq \label{Phi-C-comp-GLSM5}
\begin{align}
\phi_a
\ &= \ 
- 2 \i \, M_{c,a} 
\, , \\
D_{\Phi,a}
\ &= \ 
- \i \, D_{c,a}
+ \frac{1}{2} (\del_0 - \del_1) B_{c\+,a}
- \frac{\i}{2} (\del_0 + \del_1) A_{c=,a}
+ \frac{\i}{2} (\del_0^2 - \del_1^2) \phi_{c,a}
\, , \\
\wt{\lambda}_{\pm,a}
\ &= \ 
- \sqrt{2} \, \lambda_{c\pm,a}
\mp \i (\del_0 + \del_1) \chi_{c\mp,a}
\, ,
\end{align}
\esubeq
whilst the bosonic complex fields $F_{c,a}$, $G_{c,a}$, 
$N_{c,a}$, and the fermionic Weyl spinor fields $\psi_{c\pm,a}$, $\zeta_{c\pm,a}$ 
have no relations to the component fields of $\Phi_a$.

Next we focus on (\ref{Psi-Xi-C-GLSM5}), which gives rise to the following relations:
\bsubeq \label{Psi-Xi-C-comp-GLSM5}
\begin{gather}
\begin{alignat}{2}
r^1 \ &= \ 
- g^2 y^1
+ g^2 \sum_{a=1}^k (\phi_{c,a} + \ol{\phi}{}_{c,a})
\, , &\ls
\chi_{\pm} \ &= \ 
- g^2 \ol{\xi}{}_{\pm}
+ \sqrt{2} \, g^2 \sum_{a=1}^k \big( \psi_{c\pm,a} + \ol{\chi}{}_{c\pm,a} \big)
\, , \\
G \ &= \ 
\sqrt{2} \, g^2 \sum_{a=1}^k \big( F_{c,a} + \ol{M}{}_{c,a} \big)
\, , &\ls
0 \ &= \ 
- G_{\Xi} + \sqrt{2} \sum_{a=1}^k \big( G_{c,a} + \ol{N}{}_{c,a} \big)
\, , 
\end{alignat}
\\
\begin{align}
(\del_0 + \del_1) r^2 \ &= \ 
- g^2 (\del_0 + \del_1) y^2 
+ g^2 \sum_{a=1}^k \big( B_{c\+,a} + \ol{B}{}_{c\+,a} \big)
\, , \\
(\del_0 - \del_1) r^2 \ &= \ 
+ g^2 (\del_0 - \del_1) y^2 
+ g^2 \sum_{a=1}^k \big( A_{c=,a} + \ol{A}{}_{c=,a} \big)
\, .
\end{align}
\end{gather}
\esubeq
We should notice that the scalar fields $r^2$ and $y^2$ are related only with derivatives.


Let us evaluate the structure of supersymmetric vacua and the low energy effective theory of the Lagrangian (\ref{multi-GLSM54}). 
In order to investigate them, 
we describe the system in terms of the dynamical component fields.
Analyzing the equations of motion for auxiliary fields
under the relations (\ref{Phi-C-comp-GLSM5}),
we obtain the bosonic Lagrangian of (\ref{multi-GLSM54}) in the following form:
\begin{align}
\Scr{L}_{3\text{b}}
\ &= \ 
\sum_{a=1}^k \frac{1}{e_a^2} \Big\{
\half (F_{01,a})^2 
- |\del_m \sigma_a|^2
- 4 |\del_m M_{c,a}|^2
\Big\}
\nn \\
\ & \ \ \ \ 
- \frac{1}{2 g^2} \Big\{ (\del_m r^1)^2 + (\del_m r^3)^2 \Big\}
- \frac{g^2}{2} \Big\{ (\del_m y^2)^2 + (D_m \gamma^4)^2 \Big\}
\nn \\
\ & \ \ \ \ 
- \sum_{a=1}^k \Big\{ 
|D_m q_a|^2
+ |D_m \wt{q}_a|^2
\Big\}
- \sqrt{2} \, \ve^{mn} \sum_{a=1}^k \del_m \big( (\vartheta - t_{2,a}) A_{n,a} \big)
\nn \\
\ & \ \ \ \ 
- 2 g^2 \sum_{a,b=1}^k \big( \sigma_a \ol{\sigma}{}_b + 4 M_{c,a} \ol{M}{}_{c,b} \big)
- 2 \sum_{a=1}^k \big( |\sigma_a|^2 + 4 |M_{c,a}|^2 \big) 
\big( |q_a|^2 + |\wt{q}_a|^2 \big)
\nn \\
\ & \ \ \ \ 
- \sum_{a=1}^k \frac{e_a^2}{2} \big( |q_a|^2 - |\wt{q}_a|^2 - \sqrt{2} \, (r^3 - t_{1,a}) \big)^2
- \sum_{a=1}^k e_a^2 \big| \sqrt{2} \, q_a \wt{q}_a - \big( (r^1 - s_{1,a}) + \i (r^2 - s_{2,a}) \big) \big|^2
\nn \\
\ & \ \ \ \ 
+ \frac{g^2}{2} \sum_{a,b=1}^k 
(A_{c=,a} + \ol{A}{}_{c=,a}) (B_{c\+,b} + \ol{B}{}_{c\+,b})
\, . \label{multi-GLSM54-bosons3}
\end{align}
Note that the covariant derivatives are defined as
\bsubeq \label{cov-deriv-GLSM5}
\begin{gather}
D_m q_a \ = \ \del_m q_a - \i A_{m,a} \, q_a
\, , \ls
D_m \wt{q}_a \ = \ \del_m \wt{q}_a + \i A_{m,a} \, \wt{q}_a
\, , \\
D_m \gamma^4 \ = \ \del_m \gamma^4 + \sqrt{2} \sum_{a=1}^k A_{m,a}
\, .
\end{gather}
\esubeq
The Lagrangian (\ref{multi-GLSM54-bosons3}) still possesses the term of the auxiliary fields $A_{c=,a}$ and $B_{c\+,a}$.
However, this plays the crucial role in the derivation of the truly T-dualized sigma model.

\subsection{Low energy limit}

Here let us analyze the system in the IR limit $e_a \to \infty$.
The kinetic terms of $A_{m,a}$, $\sigma_a$ and $M_{c,a}$ in (\ref{multi-GLSM54-bosons3}) are frozen and they become auxiliary fields.
The potential terms for $q_a$, $\wt{q}_a$ and $r^i$ provide constraints among the fields in the IR regime.
Thus a supersymmetric vacuum can be obtained under the following condition:
\bsubeq \label{const-GLSM54}
\begin{gather}
0 \ = \ \frac{g^2}{2} \sum_{a,b=1}^k 
(A_{c=,a} + \ol{A}{}_{c=,a}) (B_{c\+,b} + \ol{B}{}_{c\+,b})
\, , \\
\sigma_a \ = \ 0 \ = \ M_{c,a}
\, , \\
|q_a|^2 - |\wt{q}_a|^2 \ = \ \sqrt{2} \, (r^3 - t_{1,a})
\, , \ls
\sqrt{2} \, q_a \wt{q}_a \ = \ 
(r^1 - s_{1,a}) + \i (r^2 - s_{2,a})
\, .
\end{gather}
\esubeq
The first condition provides the following relation between $\del_m r^2$ and $\del_m y^2$ via the constraint (\ref{Psi-Xi-C-comp-GLSM5}):
\begin{align}
0 \ &= \ \frac{g^2}{2} \sum_{a,b=1}^k 
(A_{c=,a} + \ol{A}{}_{c=,a}) (B_{c\+,b} + \ol{B}{}_{c\+,b})
\nn \\
\ &= \ 
- \frac{1}{2 g^2} (\del_m r^2)^2
+ \frac{g^2}{2} (\del_m y^2)^2 
+ \ve^{mn} (\del_m r^2) (\del_n y^2)
\, . \label{AB-const-GLSM54}
\end{align}
Indeed the original field $r^2$ becomes a function of the dual field $y^2$.
However, plugging this into (\ref{multi-GLSM54-bosons3}),
the kinetic term of $y^2$ disappears, whilst 
the (original) kinetic term of $r^2$ revives in the system.
This looks strange. 
But we should keep in mind that the term $\ve^{mn} (\del_m r^2) (\del_n y^2)$ emerges.
It is anticipated that the dynamics of $r^2$ should be finally replaced to the dynamics of $y^2$.
This phenomenon will be discussed later.
In addition, we should also remark that this condition is generated by the term $\mr{\Psi} - \ol{\mr{\Psi}}$ in (\ref{multi-GLSM54}).
The second line of (\ref{const-GLSM54}) denotes we choose the Higgs branch of the system.
The third line restricts the configuration of $(q_a, \wt{q}_a)$ to \cite{Harvey:2005ab}
\bsubeq
\begin{gather}
q_a \ = \ 
- \frac{\i}{2^{1/4}} \, \e^{- \i \alpha_a} \sqrt{R_a + (r^3 - t_{1,a})}
\, , \ls
\wt{q}_a \ = \ 
\frac{\i}{2^{1/4}} \, \e^{+ \i \alpha_a} \frac{(r^1 - s_{1,a}) + \i (r^2 - s_{2,a})}{\sqrt{R_a + (r^3 - t_{1,a})}}
\, , \\
R_a^2 \ = \ 
(r^1 - s_{1,a})^2 + (r^2 - s_{2,a})^2 + (r^3 - t_{1,a})^2 
\, .
\end{gather}
\esubeq
Again, the dependence of $r^2$ in $R_a$ is originated from the term $\mr{\Psi} - \ol{\mr{\Psi}}$ in (\ref{multi-GLSM54}).
This dependence is inevitable to generate the target space B-field in the IR limit.
Then the kinetic term of $(q_a, \wt{q}_a)$ is written in terms of $r^i$ and $A_{m,a}$ in the following way:
\begin{align}
- |D_m q_a|^2 
- |D_m \wt{q}_a|^2
\ &= \ 
- \frac{1}{2 \sqrt{2} R_a} \Big\{
(\del_m r^1)^2 + (\del_m r^2)^2 + (\del_m r^3)^2 
\Big\}
\nn \\
\ & \ \ \ \ 
- \sqrt{2} R_a \Big( \del_m \alpha_a + A_{m,a} - \frac{1}{\sqrt{2}} {\omega}_{i,a} \del_m r^i \Big)^2
\, . \label{DqDq-GLSM54}
\end{align}
The explicit form of ${\omega}_i$ is
\begin{align}
{\omega}_i \ &= \ 
\sum_{a=1}^k {\omega}_{i,a} 
\, , \ls
{\omega}_{i,a} \del_m {r}^i \ = \ 
\frac{- (r^1 - s_{1,a}) \del_m r^2 + (r^2 - s_{2,a}) \del_m r^1}{\sqrt{2} R_a (R_a + (r^3 - t_{1,a}))}
\, . \label{Omega-3}
\end{align}
Substituting (\ref{DqDq-GLSM54}) into the Lagrangian (\ref{multi-GLSM54-bosons3}), 
we can integrate out the gauge fields and we find
\bsubeq
\begin{gather}
A_{m,a}
\ = \ 
\frac{1}{2 R_a H} \Big( \del_m \wt{\vartheta} - {\omega}_i \del_m {r}^i \Big)
- \del_m \alpha_a
+ \frac{1}{\sqrt{2}} {\omega}_{i,a} \del_m {r}^i
\, , \label{gauge2-GLSM54} \\
H \ = \ 
\frac{1}{g^2} + \sum_{a=1}^k \frac{1}{\sqrt{2} R_a}
\, , \ls 
\wt{\vartheta} \ = \ 
\gamma^4 + \sqrt{2} \sum_{a=1}^k \alpha_a
\, .
\end{gather}
\esubeq
Note that ${\omega}_i$ becomes the target space B-field in the H-monopoles.
Meanwhile it is the KK-vector in the KK-monopoles.
$H$ is nothing but the harmonic function which appears in the H-monopoles and the KK-monopoles \cite{Okuyama:2005gx}. 
The functions ${\omega}_i$ and $H$ are related to each other via 
\begin{align}
\nabla_i H
\ = \ 
\nabla_i \Big( \sum_{a=1}^k \frac{1}{\sqrt{2} R_a} \Big)
\ = \ 
(\nabla \times {\omega})_i
\, .
\end{align}
Plugging the solution (\ref{gauge2-GLSM54}) with the gauge-fixing condition $\alpha_a = 0$ into (\ref{multi-GLSM54-bosons3}), the Lagrangian is reduced to
\begin{align}
\Scr{L}_{3\text{b}}
\ &= \ 
- \frac{1}{2} H \Big\{ (\del_m r^1)^2 + (\del_m r^2)^2 + (\del_m r^3)^2 \Big\}
- \sqrt{2} \, \ve^{mn} \del_m ((\vartheta - t_2) {A}_{n})
+ \ve^{mn} (\del_m r^2) (\del_n y^2) 
\nn \\
\ & \ \ \ \ 
- \frac{1}{2} H^{-1} (\del_m \wt{\vartheta})^2 
- \half ({\omega}_{2})^2 H^{-1} (\del_m r^2)^2
+ {\omega}_{2} H^{-1} 
(\del_m \wt{\vartheta}) (\del^m r^2) 
\nn \\
\ & \ \ \ \ 
- \half ({\omega}_{1})^2 H^{-1} (\del_m r^1)^2
- {\omega}_{1} {\omega}_{2} H^{-1} (\del_m r^1) (\del^m r^2)
+ {\omega}_{1} H^{-1} (\del_m \wt{\vartheta}) (\del^m r^1)  
\, . \label{multi-GLSM54-bosons4}
\end{align}
Here we used $A_n = \sum_{a=1}^k A_{n,a}$ and $t_2 A_n = \sum_{a=1}^k t_{2,a} A_{n,a}$. 
Notice that the field $\vartheta$ does exist in the topological term.

The target space geometry of the nonlinear sigma model (\ref{multi-GLSM54-bosons4})
represents the geometry of the five-branes of codimension three.
Now we compactify the $r^2$-direction on $S^1$ with radius ${\cal R}_2$.
The positions of the five-branes in the $r^2$-direction become periodic
\begin{align}
s_{2,a} \ = \ 2 \pi {\cal R}_2 \, a
\, , \ls {a} \in {\mathbb Z}
\, ,
\end{align}
and the number of images of the branes is infinite $k \to \infty$.
The positions of the branes in $r^1$ and $r^3$ (and $\vartheta$) are set to be origin $t_{1,a} = t_{2,a} = s_{1,a} = 0$.
In the ${\cal R}_2 \to 0$ limit,
the discrete sum over $a$ is approximated by the continuous integral of $a$.
Then 
we find
\bsubeq \label{Omega-kinfty-GLSM54}
\begin{gather}
H \ \xrightarrow{k\to\infty} \ 
h_0 + \sigma \log \frac{\mu}{\varrho}
\, , \ls
\sigma \ = \ \frac{1}{\sqrt{2} \, \pi {\cal R}_2}
\\
\omega_1 \ \xrightarrow{k\to\infty} \ 0
\, , \ls
\omega_2 \ \xrightarrow{k\to\infty} \ 
\omega_{\varrho} 
\ = \ 
\sigma \arctan \Big( \frac{r^3}{r^1} \Big)
\, ,
\end{gather}
\esubeq
where $\varrho^2 = (r^1)^2 + (r^3)^2$.
Notice that the IR divergence has been regularized by the renormalization scale $\mu$, and $h_0$ is the ``bare'' quantity which diverges in the IR limit.
We stress that it is difficult to introduce the renormalization scale $\mu$ in the GLSM Lagrangian.
This is the primary reason that we started from the GLSM for multi-centered five-branes of codimension {\it three}.
In this process the $r^2$-dependence of the functions $H$ and $\omega_i$ disappears.
As a result, the $r^2$-direction is smeared and we obtain the codimension two defect brane geometry.
Applying this limit, the Lagrangian is simplified to
\begin{align}
\Scr{L}_{3\text{b}}
\ &= \ 
- \frac{1}{2} H \Big\{ (\del_m r^1)^2 + (\del_m r^3)^2 \Big\}
- \sqrt{2} \, \ve^{mn} \del_m (\vartheta {A}_{n})
\nn \\
\ & \ \ \ \ 
- \frac{1}{2} H^{-1} (\del_m \wt{\vartheta})^2 
- \half K H^{-1} (\del_m r^2)^2
+ \omega_{\varrho} H^{-1} (\del_m \wt{\vartheta}) (\del^m r^2) 
+ \ve^{mn} (\del_m r^2) (\del_n y^2) 
\, . \label{multi-GLSM54-bosons5}
\end{align}
where the function $K$ is defined as $K = H^2 + (\omega_{\varrho})^2$.
However, this is not the final form of the effective theory, 
because the non-geometric coordinate $r^2$ still exists in the Lagrangian.

There is a comment: the defect KK-monopole 
(\ref{singleKKM-smeared34}) appears when we apply the same procedure (\ref{Omega-kinfty-GLSM54}) to the nonlinear sigma model derived from (\ref{multi-GLSM52}).
In the same way, the defect H-monopole 
\cite{deBoer:2012ma} is also obtained if the limit (\ref{Omega-kinfty-GLSM54}) is applied to the nonlinear sigma model of (\ref{multi-GLSM51}).

\subsection{Final step of T-duality}

Since we performed the duality transformation (\ref{dual2-GLSM5}),
the scalar field $y^2$ should represent the physical coordinate on the target space, whilst the scalar field $r^2$ should be regarded as the T-dual coordinate. 
Therefore, we finalize the T-dual procedure by integrating out the scalar field $r^2$ from the Lagrangian (\ref{multi-GLSM54-bosons5}):
\bsubeq \label{T-dual-final}
\begin{gather}
0 \ = \ \delta \Scr{L}_{3\text{b}}
\ = \ 
\del^m \Big\{ K H^{-1} \, \del_m r^2
- \omega_{\varrho} H^{-1} \, \del_m \wt{\vartheta}
- \ve_{mn} \, \del^n y^2 \Big\} \delta r^2
\, , \\
\therefore \ \ \ 
\del_m r^2 \ = \ 
H K^{-1}
\Big\{ \omega_{\varrho} H^{-1} (\del_m \wt{\vartheta})
+ \ve_{mn} (\del^n y^2) \Big\}
+ {\text{(constant vector)}}
\, .
\end{gather}
\esubeq
Due to the Lorentz invariance in the two-dimensional worldsheet, 
the constant vector has to vanish. 
Since we have integrated out the gauge field $A_{m,a}$ given by (\ref{gauge2-GLSM54}),
the field $r^2$ now involves the field $\wt{\vartheta}$ in addition to the field $y^2$.
Substituting this solution into (\ref{multi-GLSM54-bosons5}), 
we obtain the final form of the nonlinear sigma model:
\begin{align}
\Scr{L}_{3\text{b}}
\ &= \ 
- \frac{1}{2} H \Big\{ (\del_m r^1)^2 + (\del_m r^3)^2 \Big\}
- \half H K^{-1} \Big\{ (\del_m y^2)^2 + (\del_m \wt{\vartheta})^2 \Big\}
\nn \\
\ & \ \ \ \ 
- \omega_{\varrho} K^{-1} \, \ve^{mn} (\del_m y^2) (\del_n \wt{\vartheta})
- \sqrt{2} \, \ve^{mn} \del_m (\vartheta {A}_{n})
\, . \label{GLSM54-NLSM522}
\end{align}
Here $\omega_{\varrho}$ should be described in terms of the dualized radius $\wt{\cal R}_2 = \alpha'/{\cal R}_2$.
The target space geometry represents the $5_2^2$-brane with B-field (\ref{single522}).
We stress that the presence of the term $(\mr{\Psi} - \ol{\mr{\Psi}})$ in the GLSM (\ref{multi-GLSM54}) is inevitable to realize the correct T-duality \cite{Buscher:1987sk}, even though this involves the non-derivative terms of $r^2$.

\section{Conclusion and discussions}
\label{sect-conclusion}

In this paper, we constructed the gauged linear sigma model (GLSM) for
multi-centered five-branes of codimension three, which is promoted to the exotic five-brane.
We began with the two-dimensional $\N = (4,4)$ supersymmetric $U(1)^k$ gauge theory $\Scr{L}_1$ representing the multi-centered $k$ H-monopoles.
The position moduli of the H-monopoles are 
introduced as the FI parameters $(s_a, t_a)$. 
We then T-dualized the model to that for the multi-centered $k$ KK-monopoles $\Scr{L}_2$.
The dualization is performed through the introduction of the auxiliary
superfields $B$ and $\Gamma$. 
On the other hand, as a way to realize T-duality along another
direction, we introduced the auxiliary superfields $R$ and $\Xi$ in
$\Scr{L}_2$ and found the new GLSM
$\Scr{L}_3$ which is expected to describe the
``$5_2^2$-brane before smearing''. 
Furthermore, 
the introduction of the superfields $R$ and $\Xi$ in $\Scr{L}_1$ allows us to
find another GLSM $\wt{\Scr{L}}_2$ which would govern the geometry of a five-brane.
The relation among these models are found in Table 1. 
\begin{center}
\begin{tabular}{c}
\slb{.85}{\fbox{
\begin{tabular}{ccc}
GLSM for H-monopoles $\Scr{L}_1$ (\ref{multi-GLSM51}) 
& $\xrightarrow{\ \ \text{T-dual $\Theta \to \Gamma$} \ \ }$
& GLSM for KK-monopoles $\Scr{L}_2$ (\ref{multi-GLSM52}) 
\\
$\downarrow$ & & $\downarrow$
\\
\slb{.75}{\renewcommand{\arraystretch}{.8}
\begin{tabular}{c}
T-dual $\Psi \to \Xi$ 
\end{tabular}
} & 
\ovalbox{\renewcommand{\arraystretch}{.8}
\begin{tabular}{c}
defect five-branes \cr
($k \to \infty$ limit)
\end{tabular}
} 
& 
\slb{.75}{\renewcommand{\arraystretch}{.8}
\begin{tabular}{c}
T-dual $\Psi \to \Xi$ 
\end{tabular}
}
\\
$\downarrow$ & & $\downarrow$
\\
GLSM for a defect five-brane $\wt{\Scr{L}}_2$ (\ref{multi-GLSM53})
& $\xrightarrow{\ \ \text{T-dual $\Theta \to \Gamma$} \ \ }$
& GLSM for $5_2^2$-brane $\Scr{L}_3$ (\ref{multi-GLSM54})
\end{tabular}
}}
\\
{Table 1: GLSMs for (defect) five-branes.}
\end{tabular}
\end{center}

We then examined the low-energy limit of the GLSM $\Scr{L}_3$. 
The potential terms provide constraints on the component fields in the IR limit.
The gauge fields $A_{m,a}$ become auxiliary fields and are integrated out.
The resulting nonlinear sigma model represents codimension three
five-branes.
In order to find the codimension two $5_2^2$-brane geometry, 
we compactify the $r^2$-direction on $S^1$ and collect together 
all the periodic array of $k \to \infty$ images of the codimension
three brane.
In the limit of the small compactification radius, the
discrete sum $\sum_{a=1}^{\infty}$ over the FI parameters $s_{2,a}$ 
is approximated by the integral over the position modulus $s$.
This smearing procedure results in codimension two (defect) branes.
Finally, we integrated out the non-geometric coordinate
$r^2$ and obtained the nonlinear sigma model for the exotic
$5_2^2$-brane (\ref{GLSM54-NLSM522}).

Some comments are in order. 
Even though the isometry along the T-dual circle is
absent, we can perform the dualization in the language of the GLSMs.
It is also worthwhile to emphasize that 
the smearing procedure along the compact direction 
is incorporated in the GLSM Lagrangians 
which allow us to deal with the codimension two defect branes.

Despite that the GLSM (\ref{multi-GLSM54}) governs the five-branes of codimension three, 
we stress that the model involves {\it quantum} aspects of the $5_2^2$-brane of codimension two.
Let us recall the relation between an NS5-brane on $S^1$ and 
a KK-monopole. The NS5-brane accompanies an
infinite number of its image
array in the compact direction. 
We know that the Callan-Harvey-Strominger solution \cite{Callan:1991dj} of
the codimension four NS5-brane becomes that of the codimension three H-monopole
(smeared NS5-brane) after performing the continuous sum over the images in the small
radius limit of $S^1$.
This smeared NS5-brane and the KK-monopole solutions are related by T-duality.
On the other hand, when one performs the discrete sum over the images of
the NS5-brane in the {\it finite} radius, one obtains the localized NS5-brane
 and the harmonic function $H$ of the solution depends on the $S^1$ coordinate
$\vartheta$ \cite{Gregory:1997te}.
Consequently the isometry along the $\vartheta$-direction
is broken.
From the perspective of the worldsheet theory, 
it is discussed that this localization in $S^1$ is caused by the
instanton effect of the GLSM for the H-monopole \cite{Tong:2002rq}.
Similarly, the instanton effect of the GLSM for the
KK-monopole causes the localization of the non-geometric winding coordinate
$\vartheta$ on the T-dualized picture \cite{Harvey:2005ab, Okuyama:2005gx}.

Now we consider the relation between the KK-monopole and the
$5_2^2$-brane. 
The situation is quite parallel to the H-monopoles and the
KK-monopoles
of codimension {\it three}. 
We have approximated the discrete sum over all images of the ``$5_2^2$-brane
before smearing'' by the continuous integral in the small $S^1$-radius limit.
In this process, the $r^2$-dependence of the harmonic function $H$ disappeared and
the $5_2^2$-brane, 
the exotic object of codimension {\it two}, was obtained.
Analogous to the H- and KK-monopoles, the localization in the $r^2$-direction 
 is recovered by keeping the $S^1$ radius finite.
At first sight, we expect that 
this localization is due to the instanton effects 
of the GLSM $\Scr{L}_3$.
However this is not correct. 
As discussed in \cite{Tong:2002rq, Harvey:2005ab}, 
the notion of the constrained instantons \cite{Affleck:1980mp}
in the limit $g \to 0$ is useful.
In this limit, one encounters the truncated model 
by setting $\sigma = M_{c} = \wt{q} = r^1 = r^2 - s = 0$ 
in the small $S^1$-radius limit of (\ref{multi-GLSM54-bosons3}). 
The resulting model is just the Abelian-Higgs model with a FI term. 
Performing the Bogomol'nyi completion, we find that 
this model accommodates the Abrikosov-Nielsen-Olesen (ANO) vortex solutions with topological term
$\vartheta F_{01}$ for constant $\vartheta$.
This observation leads to an indication that 
the $\vartheta$ ({\it not} $r^2$) corrections to the 
geometry of the nonlinear sigma model is caused in the IR limit \cite{Tong:2002rq, Harvey:2005ab, Okuyama:2005gx}.
This is a conceivable result since the 
instanton effects in the KK-monopole \cite{Harvey:2005ab} induces the $\vartheta$
dependence of the geometry but does not break the isometry along $r^2$-direction. 
Therefore the Buscher rule translates the
$\vartheta$ dependence of the KK-monopole geometry into that of the $5_2^2$-brane.
In order to elucidate this structure in the string sigma model viewpoint,
it is indispensable to scrutinize the quantum aspects of the GLSM for the exotic five-brane (\ref{multi-GLSM54}),
which will be exhibited in the next paper \cite{KimuraSasaki2}.

The $r^2$-dependence of the $5_2^2$-brane geometry may be interpreted as 
 the discrete sum over the $r^2$-direction of the KK-monopole on $S^1$. 
Although this breaks the isometry along the $r^2$-direction and the Buscher rule goes out of use, 
the doubled formalism of nonlinear sigma model \cite{Hull:2006va, Hull:2009mi, Jensen:2011jna} would help us to study
the effect of the winding coordinate $r^2$ on the $5_2^2$-brane side.
It is also interesting to study other branes such as $Q$- and $R$-branes
\cite{Hassler:2013wsa} in the worldsheet description.

We also have a comment on exotic five-branes in heterotic string theory.
In the presence of non-vanishing $H$-flux, non-geometric backgrounds would also  play a central role in heterotic string \cite{Becker:2009df, McOrist:2010jw}. 
Introducing exotic branes and applying string dualities to multiple five-branes \cite{Kimura:2009hx, Kimura:2009tb}, one might find a new interpretation of ``negative tension'' branes.

\section*{Acknowledgements}

The authors would like to thank
Yuho Sakatani and
Masaki Shigemori
for valuable discussions.
They are also grateful to 
Koji Hashimoto,
Dieter L\"{u}st,
Yutaka Matsuo,
Shuhei Sasa,
Savdeep Sethi,
Yuji Tachikawa,
Satoshi Watamura, 
Satoshi Yamaguchi,
and Naoto Yokoi
for useful comments.
The work of S.~S. is supported in part by Sasakawa Scientific Research Grant from The Japan Science Society and Kitasato University Research Grant for Young Researchers.

\begin{appendix}
\section*{Appendix}

\section{Supergravity solutions of five-branes}
\label{sect-fivebrane-sols}

Here we exhibit supergravity solutions of five-branes which are closely related to the descriptions in the main part of the current paper.

\subsection{Multi-centered KK-monopoles}

We first introduce the solution of multi-centered KK-monopoles \cite{Gross:1983hb, Sorkin:1983ns} (for conventions, see \cite{deBoer:2012ma}):
\bsubeq \label{KKM-metric}
\begin{gather}
\d s^2_{\text{KKM}} \ = \ 
\d x_{056789}^2
+ H (\vec{r}) \, \d x_{123}^2
+ H (\vec{r})^{-1} \big( \d \wt{x}^4 + {\omega} \big)^2
\, , \ls
\vec{r} \in {\mathbb R}_{123}^3
\, , \\
H (\vec{r}) \ = \ 
1 + \sum_p H_p
\, , \ls
H_p \ = \ 
\frac{\wt{\cal R}_4}{2| \vec{r} - \vec{r}_p|}
\, , \\
\d {\omega} \ = \ 
*_3 \d H
\, , \ls
\e^{2 \phi} \ = \ 1
\, , \ls
B_2^{\text{dyonic}} \ = \ 
\beta \, \d \Big\{ \frac{1}{H} \Big( \d \wt{x}^4 + {\omega} \Big)
\Big\}
\, .
\end{gather}
\esubeq
Here $\wt{\cal R}_4$ is the radius of the circle in the Taub-NUT space ${\cal M}_4$, which is the transverse geometry of the KK-monopoles.
$H (\vec{r})$ is the harmonic function. $\omega$ is the KK-vector. 
Normally, the dilaton and the B-field are trivial on this geometry in string theory.
However, one can introduce a selfdual two-form on the Taub-NUT space such as $B_2^{\text{dyonic}}$ associated with the dyonic coordinate $\beta$. 
This should be related to the forth collective coordinate of the KK-monopoles \cite{Sen:1997zb}.
In the framework of the GLSM,
one can easily capture the feature of this dyonic mode \cite{Harvey:2005ab}.

\subsection{Defect five-branes} 

It is also worth describing defect five-brane solutions. 
One typical example is obtained from the KK-monopole (\ref{KKM-metric}) by smearing along $x^2$-direction:
\bsubeq \label{singleKKM-smeared34}
\begin{gather}
\d s_{\text{KKM}}^2 \ = \ 
\d x_{056789}^2
+ H \Big\{ \d \varrho^2 + \varrho^2 (\d \vartheta_{\varrho})^2 + (\d x^2)^2 \Big\}
+ H^{-1} \big( \d \wt{x}^{4} + {\omega}_{\varrho} \big)^2
\, , \\
{\omega}_{\varrho} \ = \ - \sigma' \vartheta_{\varrho} \, \d x^2
\, , \ls
H \ = \ 
h_0 + \sigma \log \frac{\mu}{\varrho}
\, , \ls
\sigma \ = \ 
\frac{\wt{\cal R}_4}{2 \pi {\cal R}_2}
\, , \\
B_2^{\text{dyonic}} \ = \ 
\beta \, \d \Big\{ \frac{1}{H} (\d \wt{x}^4 + \omega_{\varrho} ) \Big\}
\, .
\end{gather}
\esubeq
Here ${\cal R}_2$ is the radius of the compactified direction $x^2$.
A remarkable feature is that the harmonic function is given by the logarithmic form including the renormalization scale $\mu$, while $h_0$ is the bare quantity which diverges in the IR limit.

Another example is the $5_2^2$-brane \cite{deBoer:2012ma}:
\bsubeq \label{single522}
\begin{gather}
\d s_{\text{$5_2^2$}}^2 \ = \ 
\d x_{056789}^2
+ H \, \Big\{ \d \varrho^2 + \varrho^2 (\d \vartheta_{\varrho})^2 \Big\}
+ H K^{-1} \Big\{ (\d \wt{x}^2)^2 + (\d \wt{x}^4)^2 \Big\}
\, , \\
\e^{2 \phi} \ = \ H K^{-1}
\, , \ls
B_2 \ = \ 
- (\sigma' \vartheta_{\varrho}) K^{-1} \, \d \wt{x}^2 \w \d \wt{x}^4
\, , \\
H \ = \ 
h_0 + \sigma' \, \log \frac{\mu}{\varrho}
\, , \ls
K \ = \ H^2 + (\sigma' \vartheta_{\varrho})^2
\, , \ls
\sigma' \ = \ 
\frac{\wt{\cal R}_2 \wt{\cal R}_4}{2 \pi \alpha'}
\, , \ls
\wt{\cal R}_2 \ = \ \frac{\alpha'}{{\cal R}_2}
\, .
\end{gather}
\esubeq
Here $\wt{x}^2$ is the T-dual of the original coordinate $x^2$.
$\wt{\cal R}_2$ is the dual radius of ${\cal R}_2$.
This five-brane solution is also described in terms of the logarithmic harmonic function $H$.
In addition, the metric itself depends on the coordinates $\vartheta_{\varrho}$ via the function $K$, which breaks the single-valuedness of the metric.
This represents an example of T-fold \cite{Hull:2004in} which should play the crucial role in the study of string compactifications on non-geometric flux backgrounds.

\section{Another (defect) five-brane}
\label{sect-another5}

Applying the duality transformation (\ref{dual2-GLSM5}) to the original Lagrangian (\ref{multi-GLSM51}), we obtain
\begin{align}
\wt{\Scr{L}}_2 \ &= \ 
\sum_{a=1}^k \int \d^4 \theta \, \Big\{ 
\frac{1}{e_a^2} 
\Big( - \ol{\Sigma}{}_a \Sigma_a + \ol{\Phi}{}_a \Phi_a \Big)
+ \ol{Q}{}_a \, \e^{-2 V_a} Q_a
+ \ol{\wt{Q}}{}_a \, \e^{+2 V_a} \wt{Q}_a
\Big\}
+ \int \d^4 \theta \, 
\frac{1}{g^2} 
\Big( - \ol{\Theta} \Theta \Big)
\nn \\
\ & \ \ \ \ 
+ \int \d^4 \theta \, \Big\{
- \frac{g^2}{2} \Big( \Xi + \ol{\Xi} - \sqrt{2} \sum_{a=1}^k (C_a + \ol{C}{}_a) \Big)^2
- \sqrt{2} (\mr{\Psi} - \ol{\mr{\Psi}}) \sum_{a=1}^k (C_a - \ol{C}{}_a)
\Big\}  
\nn \\
\ & \ \ \ \ 
+ \sum_{a=1}^k \Big\{
\sqrt{2} \int \d^2 \theta \, \big( \wt{Q}_a \Phi_a Q_a + s_a \, \Phi_a \big)
+ \text{(h.c.)} 
\Big\}
+ \sum_{a=1}^k \Big\{ 
\sqrt{2} \int \d^2 \wt{\theta} \, \big( t_a - \Theta \big) \Sigma_a
+ \text{(h.c.)}
\Big\}
\, . \label{multi-GLSM53}
\end{align}
This system also involves the condition (\ref{AB-const-GLSM54}) which induces the duality transformation (\ref{T-dual-final}).
In the large $k$ limit, 
the target space geometry of the nonlinear sigma model denotes another defect five-brane.
This differs from the defect KK-monopole (\ref{singleKKM-smeared34}).

\end{appendix}

}
\end{document}